\documentclass[a4paper]{article}

\usepackage{atmohead2013}
\usepackage[english]{babel}

\title{All Sky Camera instrument for night sky monitoring }

\shorttitle{All Sky Camera}

\authors{
Dusan Mandat$^{1}$,
Miroslav Pech$^{1}$,
Hrabovsky Miroslav$^{1}$,
Petr Schovanek$^{1}$,
Miroslav Palatka$^{1}$,
Petr Travnicek$^{1}$,
Michael Prouza$^{1}$,
Jan Ebr$^{1}$,

}

\afiliations{
$^1$ Institute of Physics of Academy of Science of the Czech Republic, Czech Republic  \\
\scriptsize{
}
}

\email{mandat@fzu.cz}

\abstract{The All Sky Camera (ASC) was developed as an universal device for a monitoring of the night sky quality and night sky background measurement. %chyba ACS
 ASC system consists of an astronomical CCD camera, a fish eye lens, a control computer and associated electronics. The measurement is carried out during astronomical twilight. The analysis results are the cloud fraction (the percentage of the sky covered by clouds), night sky brightness (in $mag/arcsec^{2}$) and light background in the field of view of the camera. The analysis of the cloud fraction is based on the astrometry (comparison to catalogue positions) of the observed stars.}

\begin{document}
\maketitle

%Begin a section.
\section{Introduction}
%prepsano cele
An important issue of the night sky monitoring for High-Energy Astroparticle Detectors (HEAD) is the interaction of the monitoring instrument with the measurement of high energy cosmic ray particles. Contrary to lidars or radars that are active devices interfering at least to some extend with the HEAD data taking, the all sky cameras together with robotic telescopes or pyrometers are passive instruments that can work completely independently on HEAD. The benefit of all sky cameras is that the device works during the whole measurement period, it monitors atmosphere inside full HEAD Field Of View (FOV) and it does not influence the observation of high energy atmospheric showers.  All sky cameras that are simple and cheap devices can be used to characterize immediate cloudiness, night sky brightness and their time variability not only at running experiments but also at candidate sites of future observatories \cite{bib:ICRCcam}. On the other hand compared e.g. to lidars that deliver full atmospheric profiles of measured quantities (cloud heights, aerosol optical depth vs height), the all sky cameras cannot provide such a complex and detailed information.   
   
This contribution focuses on basic hardware and data analysis principles related to all sky cameras. ASCs developed by Joint Laboratory of Optics of Palacky University and Institute of Physics ASCR in Olomouc are taken as an example.

% We could separate the monitoring instrument into two groups. First group are active instruments (radars, lidars) and second group are passive instruments (robotic telescope, all sky cameras, pyrometers etc).
%First group could be operate in the period of alignment, calibration or movement of the measurement instrument or outside the field of view (FOV) of the given U/HECR instrument.
%This criteria  decreases the effective usable time or region of monitoring. The benefit of the this group is better resolution of the measurement (e.q. the lid5ar gives whole atmospheric profile).
%The benefit of the second group is time independence of the measurement. The instrument could work during the whole measurement period, it does not affect the U/HECR equipment.
%These instruments produce a an integral values of the measurement usually.   

\subsection{System Description}

The basis of the system is an astronomical camera G$1-2000$ (Moravian Instruments a.s., Czech Repu\-blic, www.mii.cz), which is equipped with a CCD chip ICX$274$AL by SONY company. The quantum efficiency of the CCD chip is higher than $50 \%$  at $450$--$550$ nm and its spectral sensitivity covers the range from approximately $400$ to $900$ nm. The electronics has low read-out noise, a 16-bit analog-digital converter and a resolution of $1600\times1200$ pixels. The camera is equipped with a fish-eye varifocal lens (the field of view is $185$ degrees) and an electronic controlled iris. This setup is capable of detecting a star with visual magnitude up to 6 mag in zenith. %visual magnitude je ve visible spectral range automaticky and visible spectral range. 

\begin{figure*}[t]
\begin{center}
%vetsi obrazek bylo 0.4 a na jinem miste
\includegraphics[width=0.9\textwidth]{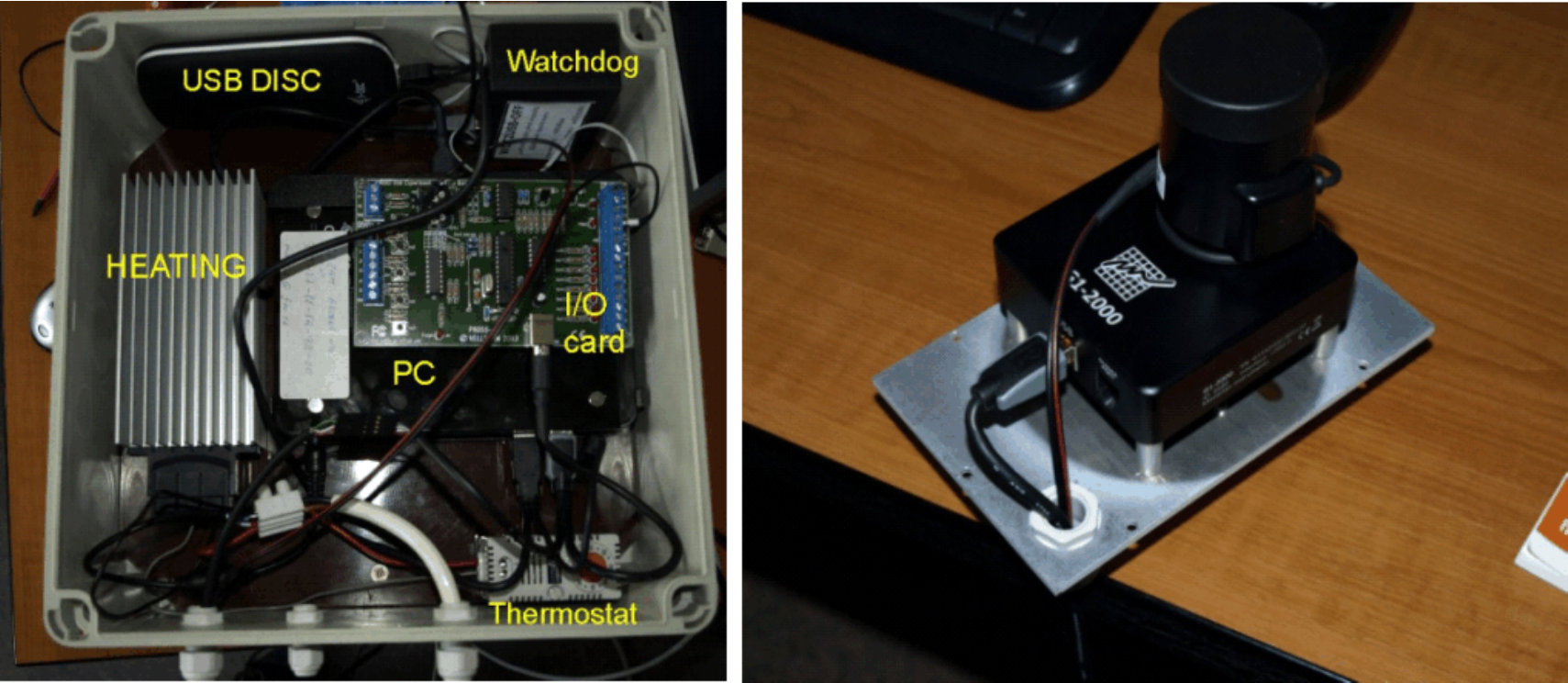}
\caption{Electronics and the inside of the camera box.}
\label{electronics}
\end{center}
\end{figure*}

A miniPC computer with USB I/O controls the system
and processes the data. The I/O card controls the iris
and the power switch of the camera.
% okopirovano z icrc, computer with serial, serial cim? A  miniPC computer with serial controls the system and processes the data. The serial $I/O$ card controls the iris and the power switch of the camera.
All electronics and a camera body (see Figure \ref{electronics}) are weatherproof. The system power could be supplied by a solar power system or local power line.
 The switching electronics turns the system ON after sunset and keeps it
going during astronomical night. The system is OFF during
the day (to save energy). Temperature of the electronics
is controlled and stabilized using internal heating system
during winter time to protect the ASC system.
% okopirovano z icrc, kde byly opraveny chyby typu aby veta mela nejed podmet ale i prisudek preklad: elektronika (fotodioda a elektronika) a vypina ... - The switching electronics (photodiode, electronics) and turns the system ON after sunset and keeps it going during astronomical night. The system is OFF during the day (to save energy). Temperature of the electronics is controlled and stabilized using internal heater during winter time to protect the ASC system.
%\\

\begin{figure*}[t]
\begin{center}
\includegraphics[width=0.43\textwidth]{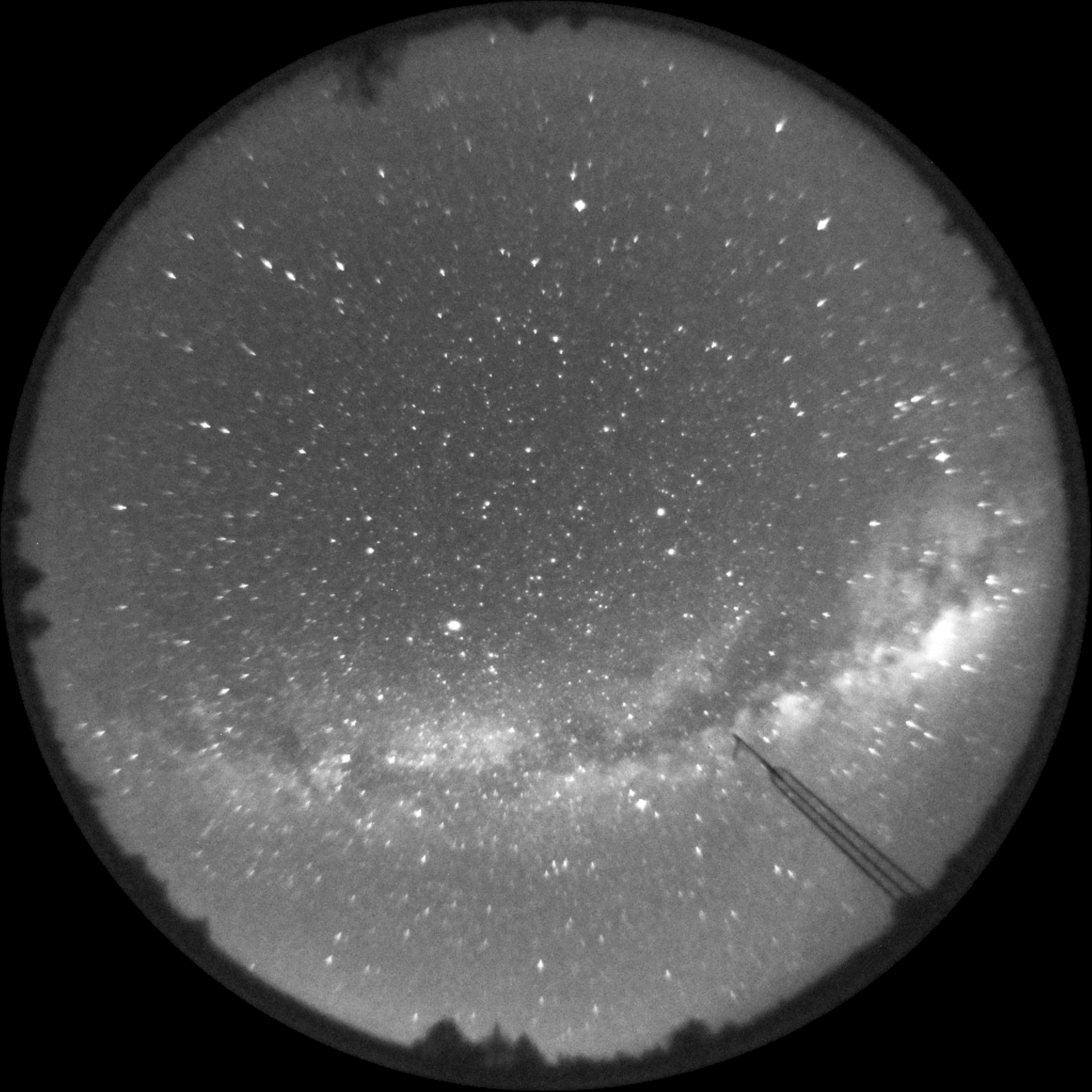}
\hspace{0.5cm}
\includegraphics[width=0.43\textwidth]{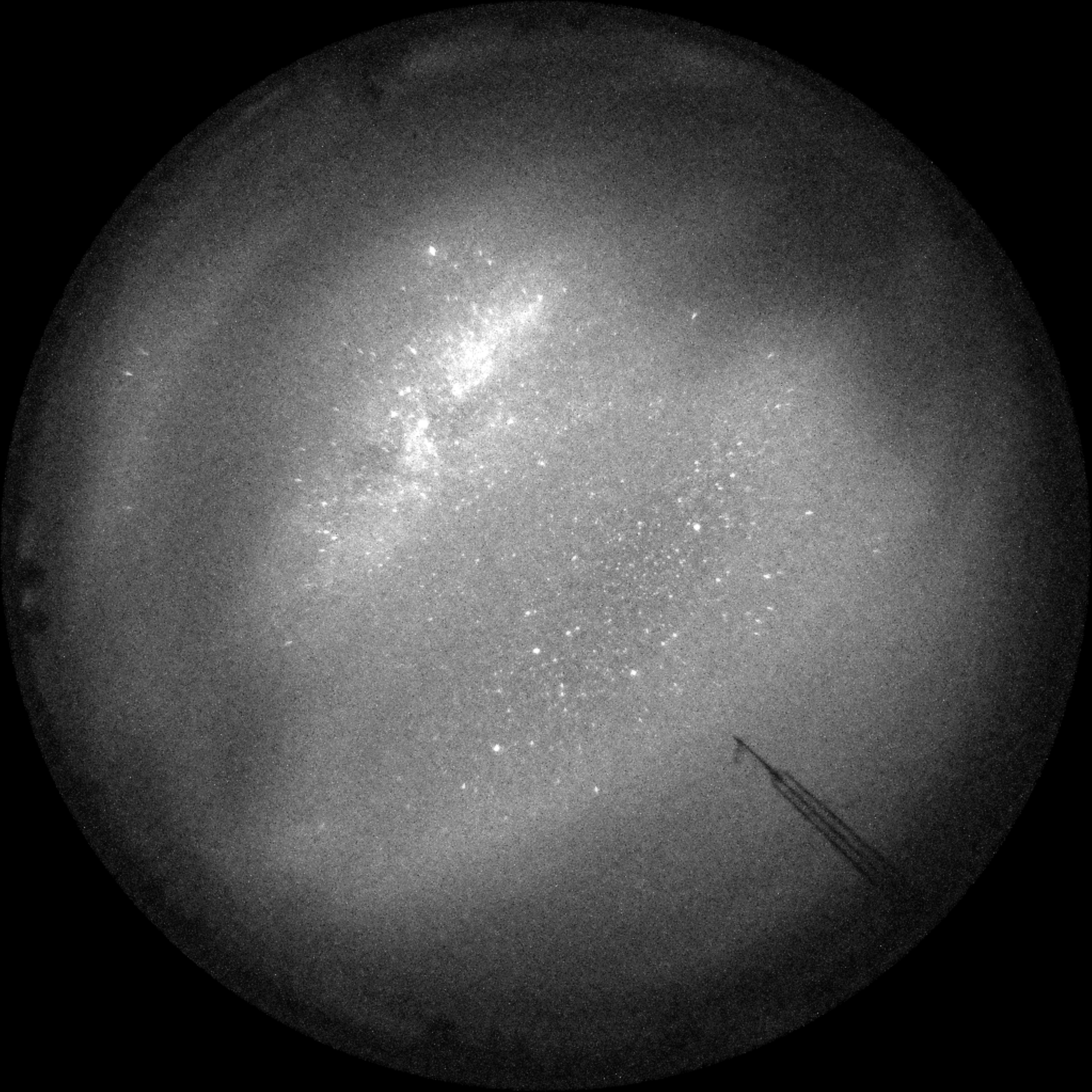}
\caption{Left: Image of clear night sky. Right: Partly cloudy night skies. Moon and clouds are visible close to horizon.}
\label{nightSky}
\end{center}
\end{figure*}

\subsection{Calibration of the field of view}

The fish-eye varifocal lens has $185^\circ$  FOV and its aberration distorts the image. The correct transformation of the incident angle to the imaged pixel position is very important.
The mechanism of the calibration is as follows:
 the light source (a spot as small as possible) 
is placed at a defined distance (at least 10 m), then the body
of camera is rotated around the axis such that the incident
light angle (corresponding to the FOV angle) varies between
0 and 90 degrees. The position of the light spot is
analyzed for each angular step. This procedure is repeated
for different plane cuts of the lens FOV. The final data
are analyzed and fitted with a polynomial function. This
calibration is regularly checked on-site by comparing the
position of detected stars with the stars in a catalog. On-site
measurements are also used to determine the variation
of the sensitivity of the system with the zenith angle.
%bylo nakopirovano icrc,  nechapu napriklad co se mysli during for image reconstruction a podobne, WE je velkym a tak dale
% WE use the light source (a spot as small as possible) at a defined distance (at least $10$ m), the body of camera is rotated around the axis such that the incident light angle (corresponding to the FOV angle) varies between $0$ and $90$ degrees. The position of the light spot is analyzed and calculated. The data are analyzed and fitted with a polynomial function. This function is used during for image reconstruction (in meaning of the optical distortion) during the operation. 
\\

\subsection{Night sky data taking and principles of analysis}
The ASC system takes full sky images  periodically in predefined moments (time period is limited by $2\times$ exposure time - night sky image and dark image).
% prepsano nechapu The ASC system takes full sky images in defined periods (limited by $2x$ exposure time - nightsky image and dark image).
 Examples of clear and partly cloudy night sky are shown in Figure \ref{nightSky}. %and \ref{nightSkyClouds}.  

%\begin{figure}[htbp]
%\begin{center}
%\includegraphics[width=0.44\textwidth]{night_sky_clouds1.pdf}
%\caption{Image of partly cloudy night sky. Moon and clouds are visible close to horizon.}
%\label{nightSkyClouds}
%\end{center}
%\end{figure}

\begin{figure}[t]
  \centering
  \includegraphics[width=0.44\textwidth,height=0.175\textheight]{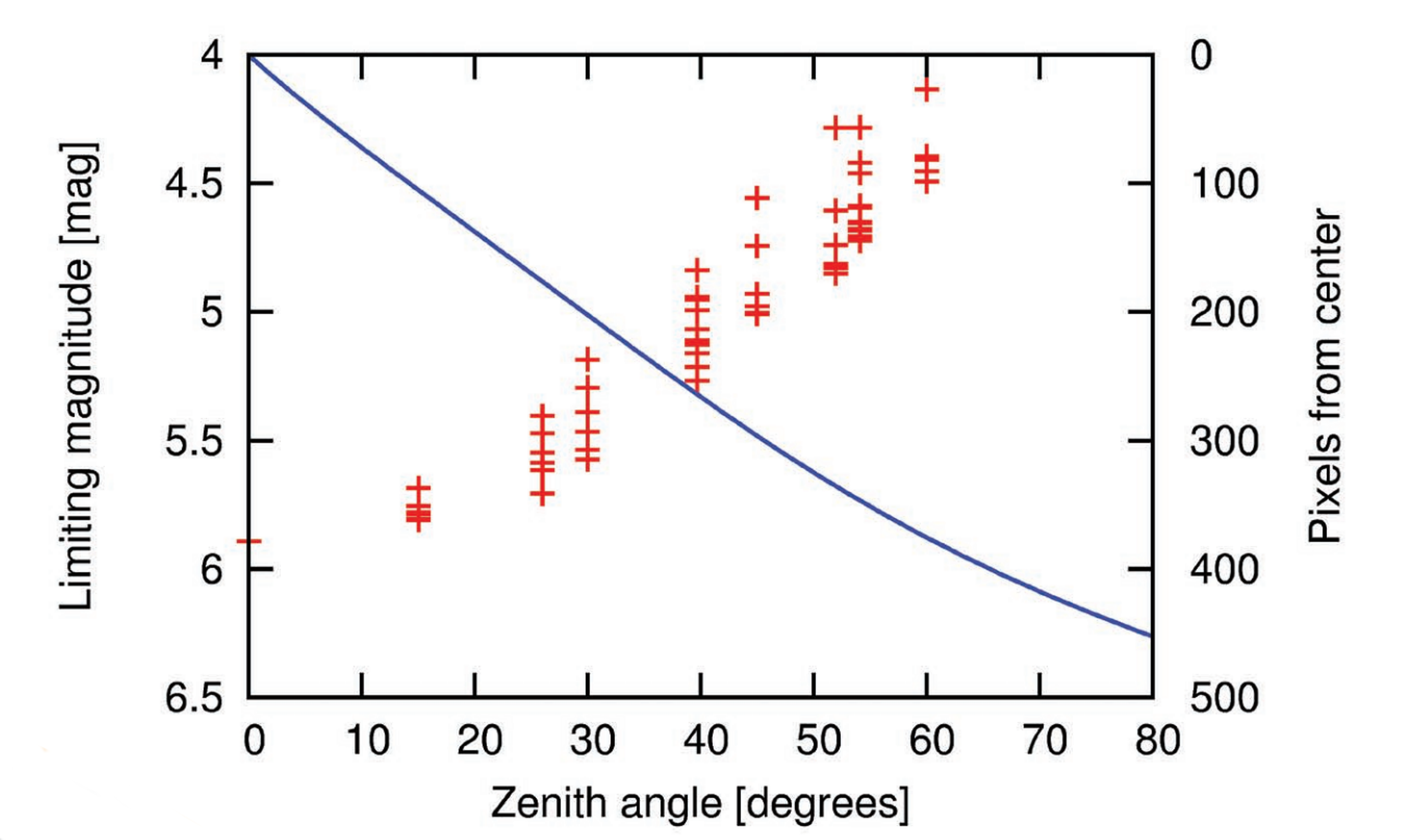}
  \caption{Zenith angle to magnitude dependence. The red points show sensitivity thresholds %measured - prah se meri?
 in different azimuthal angles as a function of the zenith angle. The blue line is the geometric calibration - distortion description. }
%caption prepsana, nesrozumitelne, jeste by to chtelo obrazek znamena 2 veci pokud to dobre chapu neni to jen zenith angle to magnitude dependence The blue line is the geometric calibration - distortion description.
  \label{zen_mag}
 \end{figure}

The measurement process consists of the following steps. 
Initially, every exposure of a night sky is followed by a dark frame
image so that subtraction of one from the other enables significant
reduction of noise in the image. The horizontal coordinates of each 
image pixel are then calculated with the calibration data and calibration
constants used to determine the horizontal coordinates, azimuth and elevation 
of each image pixel. The total charge of each pixel is proportional to the light
intensity, but the intensity of the final image varies with
respect to zenith angle due to vignetting and distortion of the
lens. The calibration of this zenith angle intensity dependence
is obtained by considering extinction in air masses, optical distortions and 
vignetting variation as a function of zenith angle. The resulting attenuation at, for
example, $60$ degrees zenith angle corresponds to $4$ mag (see Fig.~\ref{zen_mag}).

\begin{figure}[t]
  \centering
  \includegraphics[width=0.45\textwidth,height=0.26\textwidth]{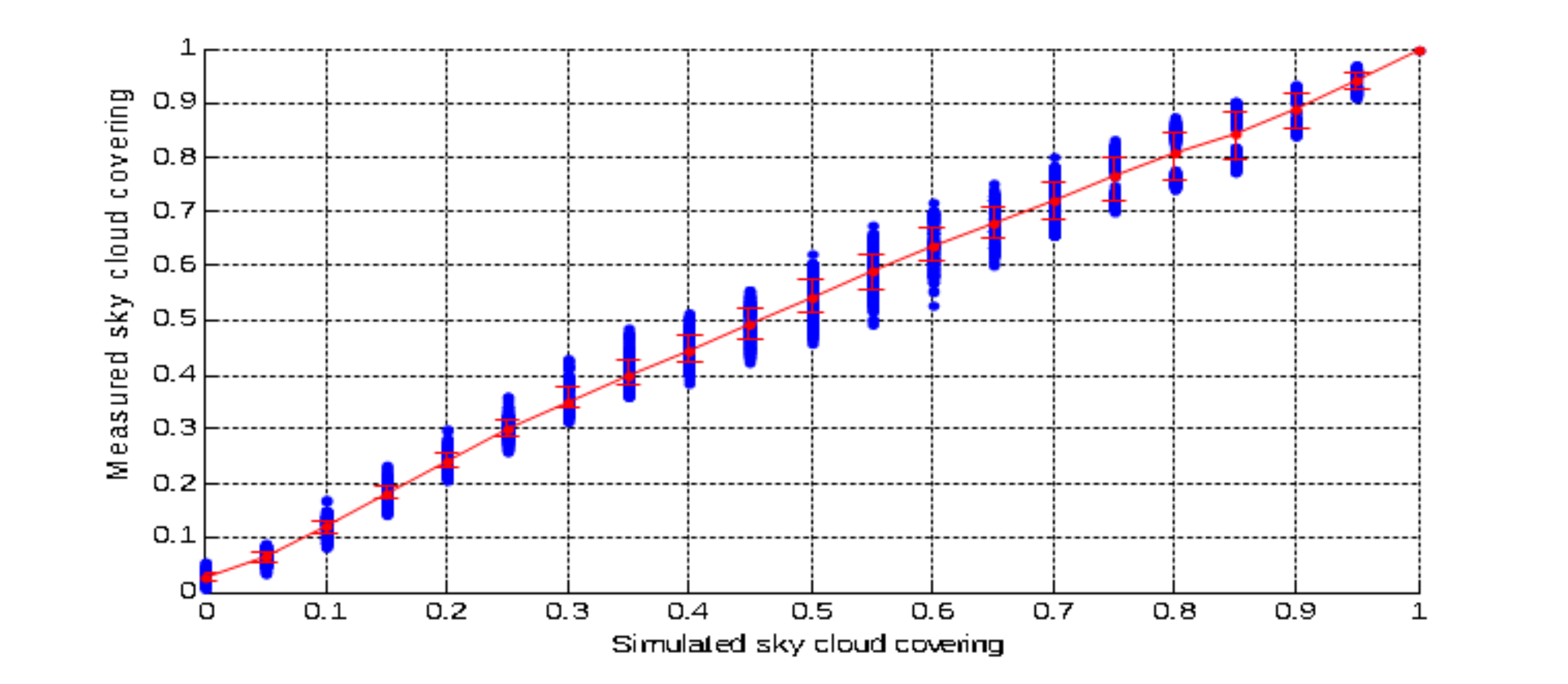}
  \caption{Error of the algorithm calculated using artificial cloud simulations with cloud fractions from $0\%$ - $100\%$ with step of $5\%$.  Horizontal axis shows the simulated cloud cover and vertical axis corresponds to values calculated by the algorithm (blue dots). The error bars (red) represent the statistical estimation of the algorithm uncertainty.}
  \label{error}
 \end{figure}

\begin{figure*}[t]
  \centering
  \includegraphics[width=\textwidth]{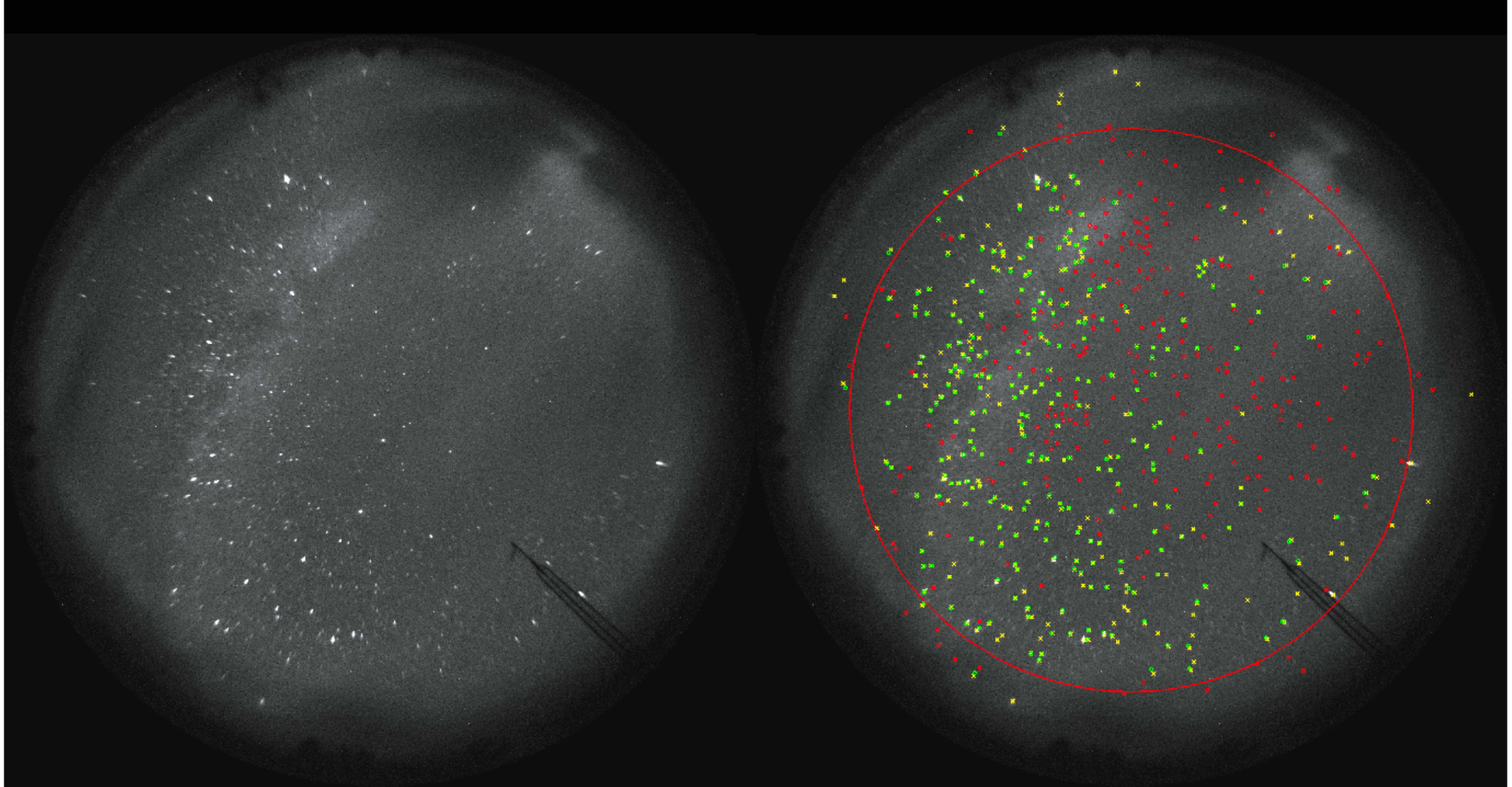}
  \caption{An example of analysis of partly cloudy night sky. The image on the left shows RAW image of the sky. The image on the right shows the analysis results, the yellow crosses indicate the detected stars, green circles catalog stars and red circles represent catalog stars without detected pairs - the region covered by cloudiness. The red circle indicates the limiting zenith angle ($60^{\circ}$).}
  \label{cloudiness_analysis}
 \end{figure*}

To estimate the cloud fraction, we investigate the presence of stars in the field of view of the camera. We separate the full sky to approximately 70 segments and analyze each segment of the sky individually to obtain the cloudiness. As a star catalog we use Yale Bright Star Catalog BSC$5$ \cite{bib:starcat}, particularly the visual magnitudes of the stars. First we estimate the position of each segment on the sky using the data from the calibration. This information is then further improved using the brightest stars from the catalog in each segment. Once the position of a segment on the sky is fixed, the stars for the segment are read from the catalog and selected according to the expected sensitivity at the given zenith angle.
 For every star in the catalog, we look for a detected star within the angular limit of $1^\circ$. If a star that fits this criteria is found, then the catalog and detected stars are flagged as paired. Figure \ref{cloudiness_analysis} shows a result from cloudiness analysis of partly cloudy night sky. The ratio of paired / unpaired stars (for all segments) gives final cloudiness of the night sky. Typically, we check about 500 catalog stars for the whole observed sky (up to $60^{\circ}$ due to high distortion close to horizon). The error of the algorithm was calculated using artificial cloud simulations (Fig.~\ref{error}) and the total uncertainties of the cloudiness
calculation are 2\% for cloudiness $<$20\% and 5\% for cloudiness $>$80\%. The cloudiness of the night sky is computed up to the maximal zenith angle with 10$^{\circ}$ steps.
%okopirovano z site reportu, predelana citace na obrazek and the total error of the cloudiness calculation is $\pm 5 \%$.  The cloudiness of the night sky could be computed up to the limiting angle with $10 ^{\circ}$ steps.     

Brightness of the night sky could be evaluated directly from a light flux. The system has been calibrated using a darkness sky measurement tool Unihedron SQM-LE \cite{bib:sqm}. The night sky was scanned overnight and recorded data both from the camera and SQM instrument were compared to each other. Subsequently the conversion formula is obtained. The data represent the brightness of the night sky in $mag/arcsec^2$ and in visible light spectrum.

\vspace{0.5cm}
\subsection{Limits of the cloudiness analysis}

\begin{figure*}[t!]
\begin{center}
\includegraphics[width=0.45\textwidth]{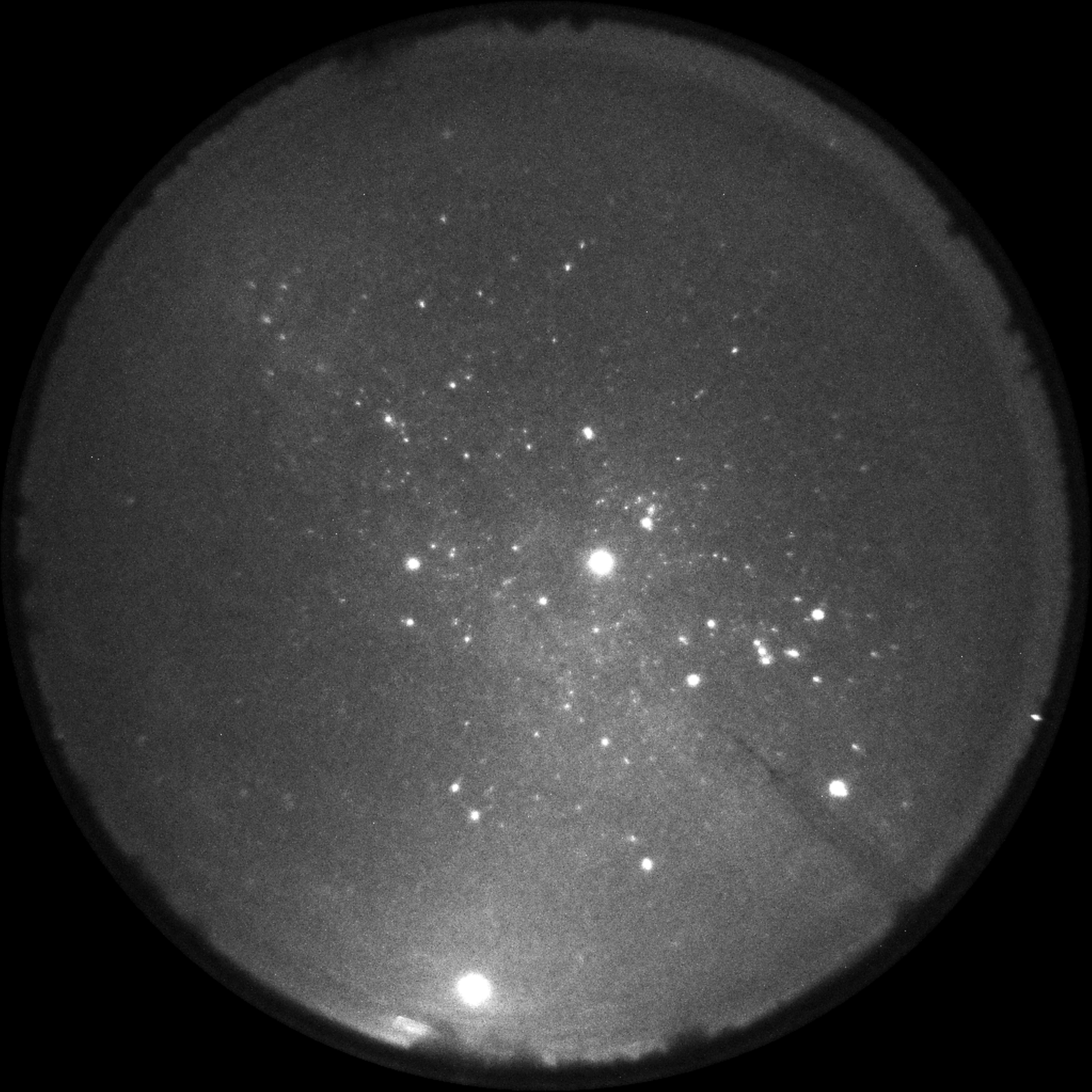}
\hspace{0.5cm}
\includegraphics[width=0.45\textwidth]{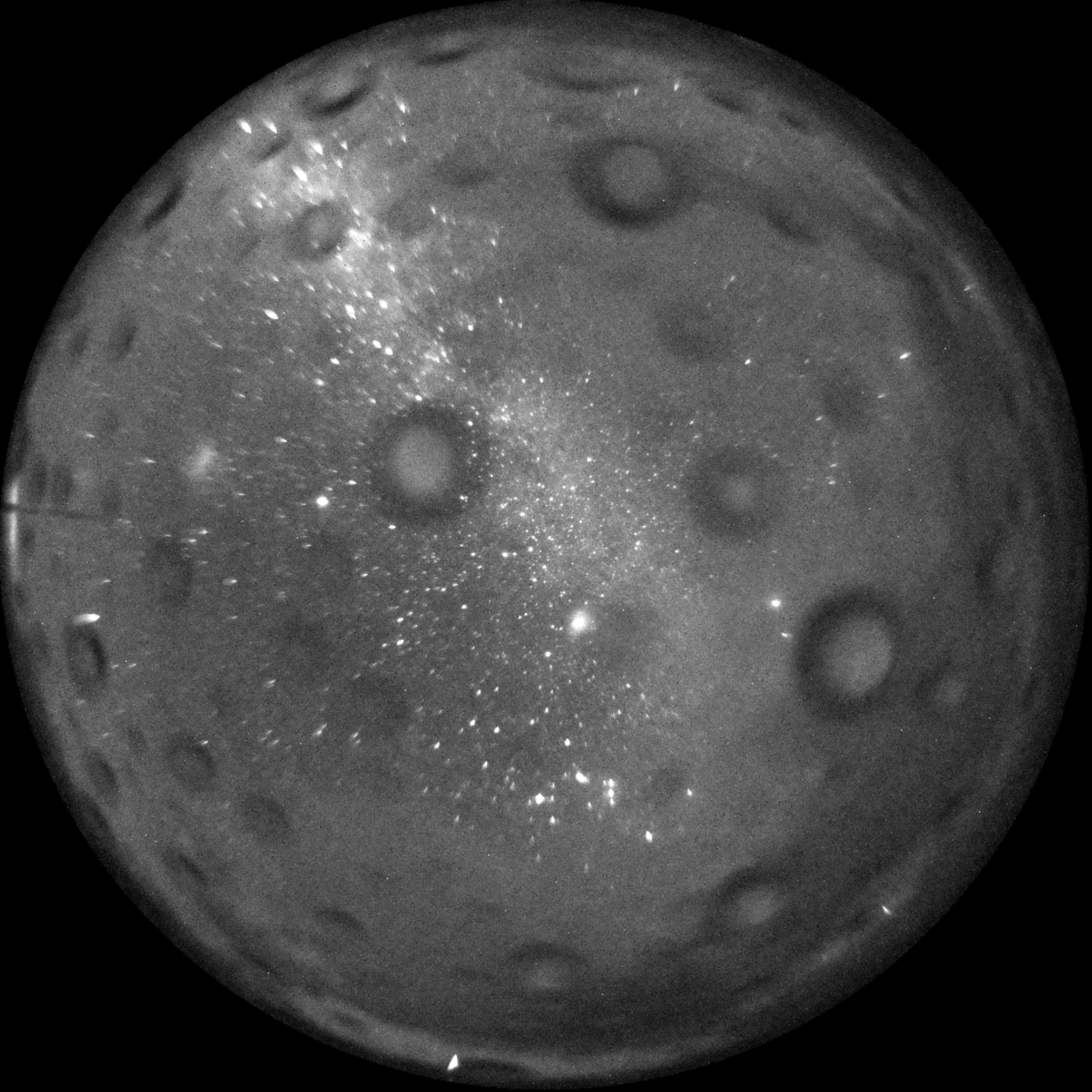}
\caption{Left: Fog or haze affects the visibility of night sky, the nearby tower is not visible (the visibility fades with the fog/haze). Right: Lens of the ASC covered with water drops. }
\label{fog}
\end{center}
\end{figure*}

The measurement using the ASC is limited by local weather conditions, camera misting and Moon position. The local weather limitations mean for example rain or snow, fog, haze and water condensation on the camera. Examples of rain, fog and snow are shown in Figure \ref{fog}.
The algorithm usually fails if the lens of the ASC is obscured in some way, for example by snow, water resulting from rain fall, haze or fog surrounding ASC, or even bird droppings from birds found locally. The presence of the Moon saturates the image and increases the background in the image, and consequently the algorithm for cloud detection could not be used at such times. 

%\subsection{File format}

%\begin{itemize}
%\item atmohead2013-XXXX.tex
%\item atmohead2013-XXXX-YY.(eps or pdf)
%\item atmohead2013-XXXX.(ps or dvi or pdf)
%\end{itemize}

%where XXXX stands for your last name (e.g. atmohead2013-chaves.pdf).

%All those files must compressed in a single package named 
%atmohead2013-XXXX.tar.gz. This is the only file that should be submitted.
 
\section{Conclusions}
%pridan text z icrc
The ASC, its application and limits are presented. The result of the analysis is the night sky cloud cover and night sky brightness. The main advantage of the all-sky camera is that it is a fully autonomous and self-reliant tool which can be used to monitor the night sky in even the world's most remote places, which happen to be just the ones most promising for astronomically related applications. 
The autonomy of the camera significantly contributes to the extremely low cost for its operation. The costs involved are concentrated mainly
in the deployment phase, namely the costs of the parts
and assembly of the camera, shipping (varies depending
on destination, the shippable weight being 10 kg) and installation (two people
working for two or three days). As demonstrated in \cite{bib:ICRCcam} the
camera can operate without maintenance for months and years after installation. However,
it is advantageous to have local support for the case of
unexpected external hindrances, but in most cases, a nonspecialist
local personnel is sufficient to fix most of the possible
issues.

%\vspace{2cm}

%caption jsem prepsal

%\begin{figure}[t!]
%\begin{center}
%\includegraphics[width=0.45\textwidth]{water1.pdf}
%\caption{Lens of the ASC covered with water drops.}
%\label{water}
%\end{center}
%\end{figure}

\vspace{2cm}

\footnotesize{{\bf Acknowledgment: }{This work was supported by the  Ministry of Education Youth and Sports of the Czech Republic under the projects LE13012, 7AMB12AR013 and LG13007}}

\end{document}